\documentclass[
  journal=largetwo,
  manuscript=article-type,
  year=2025,
  volume=XX,
]{cup-journal}

\DeclareUnicodeCharacter{22C6}{\ensuremath{\star}}
\usepackage{amsmath}
\usepackage{amssymb}
\usepackage{newunicodechar}
\usepackage{booktabs}
\usepackage{hyperref}
\usepackage[utf8]{inputenc}
\hypersetup{
    colorlinks=true,
    linkcolor=blue,
    citecolor=blue,
    filecolor=magenta,      
    urlcolor=blue,
    pdftitle={Overleaf Example},
    pdfpagemode=FullScreen,
    }

\usepackage{newunicodechar}
\newunicodechar{≃}{$\simeq$}
\newunicodechar{≲}{$\lesssim$}

\usepackage[backend=biber]{biblatex}
\title{Inferring Obscured Cosmic Black Hole Accretion History from AGN Found by JWST/MIRI CEERS Survey}

\author{Cheng-An Hsieh}
\affiliation{Department of Physics, National Taiwan University, No. 1, Sec. 4, Roosevelt Rd., Da'an Dist., Taipei City, 106216, Taiwan (R.O.C.)}
\email[Cheng-An Hsieh]{b12202042@ntu.edu.tw}

\author{Tomotsugu Goto}
\affiliation{Institute of Astronomy, National Tsing Hua University, 101, Section 2, Kuang-Fu Road, Hsinchu 30013, Taiwan (R.O.C.)}
\alsoaffiliation{Department of Physics, National Tsing Hua University, 101, Section 2, Kuang-Fu Road, Hsinchu 30013, Taiwan (R.O.C.)}

\author{Chih-Teng Ling}
\affiliation{Institute of Astronomy, National Tsing Hua University, 101, Section 2, Kuang-Fu Road, Hsinchu 30013, Taiwan (R.O.C.)}

\author{Seong Jin Kim}
\affiliation{Institute of Astronomy, National Tsing Hua University, 101, Section 2, Kuang-Fu Road, Hsinchu 30013, Taiwan (R.O.C.)}

\author{Tetsuya Hashimoto}
\affiliation{Department of Physics, National Chung Hsing University, 145, Xingda Road, Taichung 40227, Taiwan (R.O.C.)}

\author{Tom C.-C. Chien}
\affiliation{Department of Physics, National Tsing Hua University, 101, Section 2, Kuang-Fu Road, Hsinchu 30013, Taiwan (R.O.C.)}

\author{Amos Y.-A. Chen}
\affiliation{Department of Physics, National Tsing Hua University, 101, Section 2, Kuang-Fu Road, Hsinchu 30013, Taiwan (R.O.C.)}


\begin{document}

\begin{abstract}

This study presents the black hole accretion history (BHAH) of obscured active galactic nuclei (AGNs) identified from the JWST CEERS survey by Chien et al. (2024) using mid-infrared (MIR) SED fitting. We compute black hole accretion rates (BHARs) to estimate the black hole accretion density (BHAD), \(\rho_{L_{\text{disk}}}\), across \(0 < z < 4.25\). MIR luminosity functions (LFs) are also constructed for these sources, modeled with modified Schechter and double power law forms, and corresponding BHAD, \(\rho_{\text{LF}}\), is derived by integrating the LFs and multiplying by the luminosity. Both \(\rho_{\text{LF}}\) extend to luminosities as low as \(10^7 \, L_{\odot}\), two orders of magnitude fainter than pre-JWST studies. Our results show that BHAD peaks between redshifts 1 and 3, with the peak varying by method and model, z $\simeq$ 1 - 2 for \(\rho_{L_{\text{disk}}}\) and the double power law, and z $\simeq$ 2 - 3 for the modified Schechter function. A scenario where {AGN} activity peaks before cosmic star formation would challenge existing black hole formation theories, but our present study, based on early JWST observations, provides an initial exploration of this possibility. At \(z \sim 3\), \(\rho_{\text{LF}}\) appears higher than X-ray estimates, suggesting that MIR observations are more effective in detecting obscured AGNs missed by X-ray observations. However, given the overlapping error bars, this difference remains within the uncertainties and requires confirmation with larger samples. These findings highlight the potential of JWST surveys to enhance the understanding of co-evolution between galaxies and AGNs.
\end{abstract}

\section{Introduction}

\quad Active Galactic Nuclei (AGNs) are galaxies with bright cores powered by supermassive black holes (SMBHs). Understanding AGN activities is crucial for several reasons: It provides insights into star formation \citep{morganti2017many} and galaxy evolution \citep{kormendy2013coevolution}, contributes to our knowledge of SMBH growth \citep{marconi2004local}, and reveals the dynamic interaction between black holes and their host galaxies \citep{croton2006many}.

By studying black hole accretion events and black hole accretion history (BHAH), we gain a more complete understanding of how obscured AGN activity impacts galaxy evolution and star formation processes \citep{chen2013correlation, kormendy2013coevolution}. This work focuses on the co-evolution of galaxies and their central black holes.

The unified model of AGN \citep{antonucci1993unified, netzer2015revisiting} outlines their structure and explains their observed differences based on the orientation of the central black hole and surrounding dust. It categorizes AGNs into Type I and Type II, where Type I AGNs allow direct observations in optical, UV, or X-ray wavelengths. In contrast, Type II AGNs are obscured by dust, making detection challenging in these bands e.g., \citet{urry1995unified, heckman1997powerful, kauffmann2003host}. Type II AGNs can only be observed in the infrared, and there is less data available in prior studies than for Type I AGNs.

The advanced sensitivity of JWST's MIRI enables the detection of obscured AGNs that are up to 8 times fainter than those detectable by prior IR space telescopes such as AKARI \citep{goto2010evolution, goto2015evolution, goto2019infrared}, Spitzer \citep{le2005infrared}, and WISE \citep{yang2021jwst}. This extended sensitivity afforded by JWST enables the study of AGNs at higher redshifts, providing critical data on the evolution of the faint population of obscured AGNs. The more comprehensive wavelength range covers 5 to 25 $\mu$m, offering comprehensive coverage for detecting high redshift objects.

To reduce selection bias, mid-infrared (MIR) observations are especially useful for investigating obscured AGNs \citep{lacy2006optical}. Studying Type II AGNs helps reduce selection bias, leading to more accurate assessments of black hole accretion history \citep{yang2023ceers}. Previous studies also indicate a strong correlation between MIR luminosity and X-ray luminosity  \citep{hickox2018obscured}, showing the importance of MIR observations in investigating AGNs. Therefore, with its high sensitivity, JWST is crucial because it enables us to observe fainter and more distant AGN populations.

Furthermore, {by} constructing the luminosity function (LF), which describes the distribution of galaxy luminosities in a given volume, we can uncover the evolution of dusty AGNs. {Previous studies\citep[e.g.,][]{saunders199060, goto2010evolution, ling2024exploring}, using sources identified with infrared space telescopes such as IRAS  \citep{neugebauer1984infrared} and Spitzer \citep{werner2004spitzer}, have demonstrated the efficacy of measurements of the IR LF, which effectively trace the distribution of BH growth rates and constrain the BHAH.}

This study aims to build the first AGN LF for candidates identified by JWST, providing insights into BHAH.
1
This study is organized as follows: Section 2 describes our data and utilizes the AGN candidates identified by \citet{chien2024finding} to estimate the black hole accretion density (BHAD). We compare our results with previous studies that used different selection methods for AGNs \citep{yang2023ceers}. Section 3 details the construction of the infrared luminosity function. From this, we derive the luminosity density (LD), which signifies the average infrared energy emitted by the AGN within a specific volume. Through the relation between LD and BHAD, we determine the BHAD, discuss the results using a different method, and compare it to previous studies.

\section{AGN Selection and BHAD Estimation}

\subsection{AGN Selection}



The JWST Cosmic Evolution Early Release Science (CEERS, \citealt{finkelstein2017cosmic}) Survey provides observational data in the Extended Groth Strip (EGS) legacy field. It employs the Near-Infrared Camera (NIRCam), the Mid-Infrared Instrument (MIRI), and the Near-Infrared Spectrograph (NIRSpec) to capture light across a wavelength range from 0.8 $\mu$m to 21 $\mu$m. \citet{ling2024exploring} utilized MIRI observations using six broad-band filters: F770W, F1000W, F1280W, F1500W, F1800W, and F2100W, covering a continuous wavelength range of 7.7–21.0 $\mu$m. The dataset consists of four pointings with different completeness and depth, resulting in an effective sky coverage of approximately 9.2 arcmins$^2$.

\citet{chien2024finding} specifically focused on MIRI imaging data from CEERS, cross-matching it with the CANDELS-EGS multiwavelength catalog \citep{stefanon2017candels}. This process linked MIRI-detected sources with photometric and spectroscopic data from CANDELS-EGS, covering from UV to IR, using 15 + 6 photometric bands for 573 sources.

To derive the physical properties of faint AGNs, \citet{chien2024finding} performed spectral energy distribution (SED) fitting using the Code Investigating GALaxy Emission (CIGALE) v2022.1 \citep{boquien2019cigale}. The SED fitting results provided insights into the physical properties of faint, obscured AGNs, such as redshift, AGN luminosity, and AGN contribution, allowing for a more detailed analysis.


{Galaxy candidates from \citet{chien2024finding} need detection in at least three bands, unlike the criteria in \citet{yang2023ceers} which require only two bands. A 10 $\mu$m 80 per cent completeness flux limit \citep{ling2022galaxy, wu2023source} is applied to exclude faint sources. {Stars are also removed by checking the performance of SED fittings. Through manual inspection, if the SED fitting shows a significant drop in the mid-infrared, which is typical for stellar sources, the object is excluded from our sample.} Finally, 253 sources are categorized into three populations based on their AGN contributions, defined by ($f_{\text{AGN, IR}}$, \citealt{boquien2019cigale}):}

\begin{itemize}
\item \textbf{Star-forming galaxies (SFGs)}: $f_{\text{AGN, IR}} \leq 0.2$
\item \textbf{Composites}: $0.2 < f_{\text{AGN, IR}} < 0.5$
\item \textbf{AGNs}: $f_{\text{AGN, IR}} \geq 0.5$
\end{itemize}

The AGN contribution is derived by infrared luminosity based on the SED fitting:
\begin{equation}
f_{\text{AGN, IR}} = \frac{L_{\text{AGN}}}{L_{\text{TIR}}}
\end{equation}
where $L_{\text{AGN}}$ denotes the AGN luminosity and $L_{\text{TIR}}$ is the total infrared luminosity among 8-1000 $\mu$m, as defined by \citet{kennicutt1998star}. 

211 SFGs, 30 Composites, and 11 AGNs were identified, compared to \citet{yang2023ceers} with 433 SFGs, 102 Composites, and 25 AGNs. {The smaller sample size in this work is due to the stricter selection criteria adopted.}

Additionally, we check the AGN inclination ($i$) from CIGALE results, categorizing those with angles near 0, 10, 20, or 30 degrees as Type I, and those around 70, 80, or 90 degrees as Type II.

\begin{figure}
    \centering
    \includegraphics[width=0.6\textwidth]{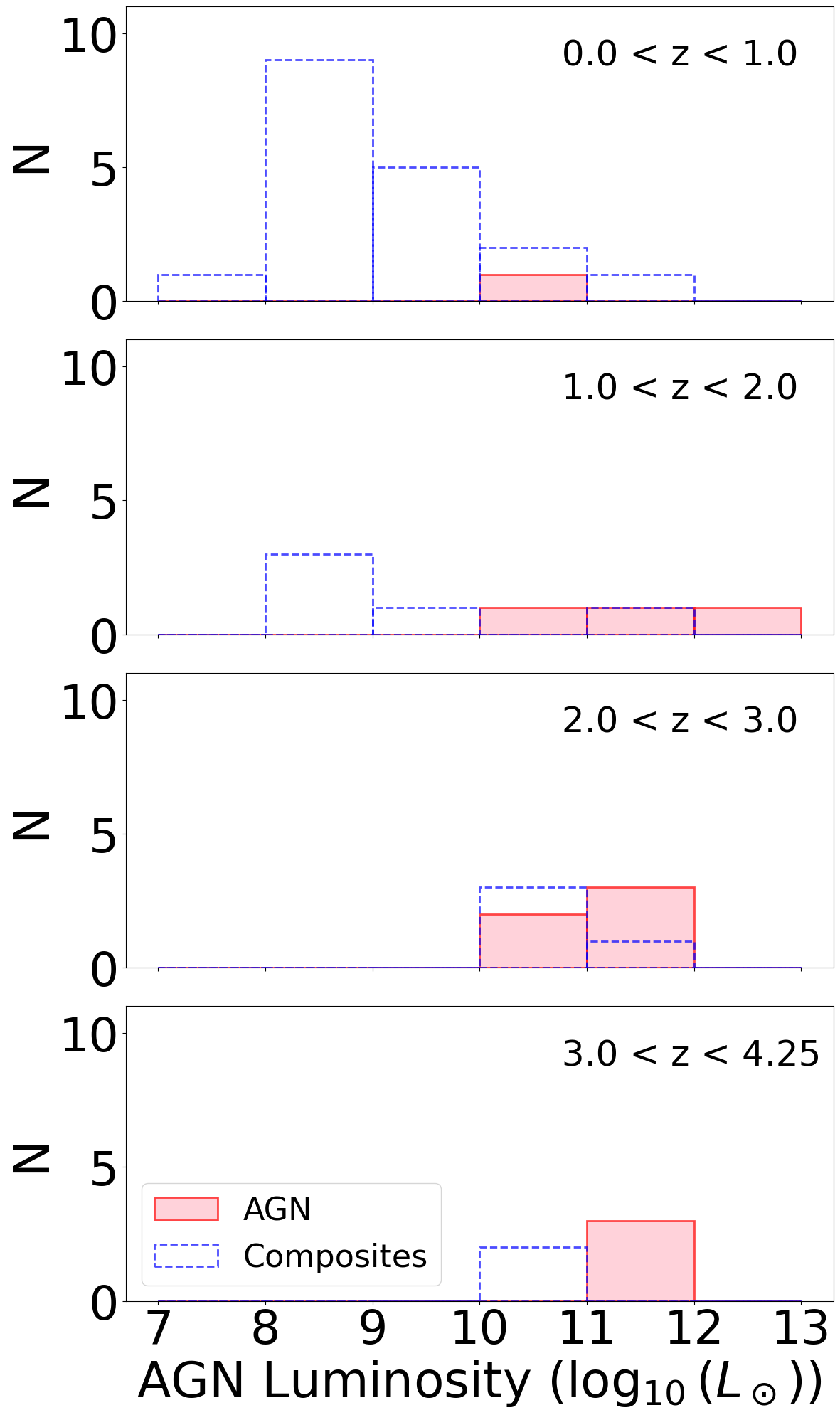}
    \caption{Luminosity histograms for each redshift bin. Composites and AGNs are represented by cyan and red, respectively. }
    \label{fagn}
\end{figure}

To infer BHAD, we take composites and AGNs (i.e., galaxies with $f_{\text{AGN, IR}} > 0.2$) as our galaxy sample.
Figure \ref{fagn} shows the luminosity distribution of our sample at each redshift bin. We divide our candidates into four redshift bins ([0, 1], [1, 2], [2, 3], [3, 4.25]) to obtain a sufficient sample size in each bin and highlight evolution features, especially at $z \simeq 1 - 2$ and $2 - 3$.
At the lowest redshift, the luminosities of the faintest AGNs are around $10^7$ to $10^8$ $L_{\odot}$, 1-2 orders of magnitude fainter than candidates from the literature for $z \lesssim 2$ (e.g. \citealt{traina2024a3cosmos}).

\subsection{BHAD Estimation}\label{2.2}
First, for comparison, we follow \citet{yang2023ceers} to obtain the black hole accretion density (BHAD) of our galaxy sample.
Their BHAD is simply derived from the total black hole accretion rate (BHAR):
\begin{equation}
\text{BHAR} = \frac{L_{\text{disk}}(1-\varepsilon)}{\varepsilon c^2}
\label{BHAR}
\end{equation}
where $L_{\text{disk}}$ is the AGN accretion disk luminosity averaged by the viewing angle, directly taken from the CIGALE parameter "accretion power" \citep{yang2018linking}. $\varepsilon$ is the radiative efficiency (set to 0.1), and $c$ is the speed of light.

\begin{figure*}
    \centering
    \includegraphics[width=0.6\textwidth]{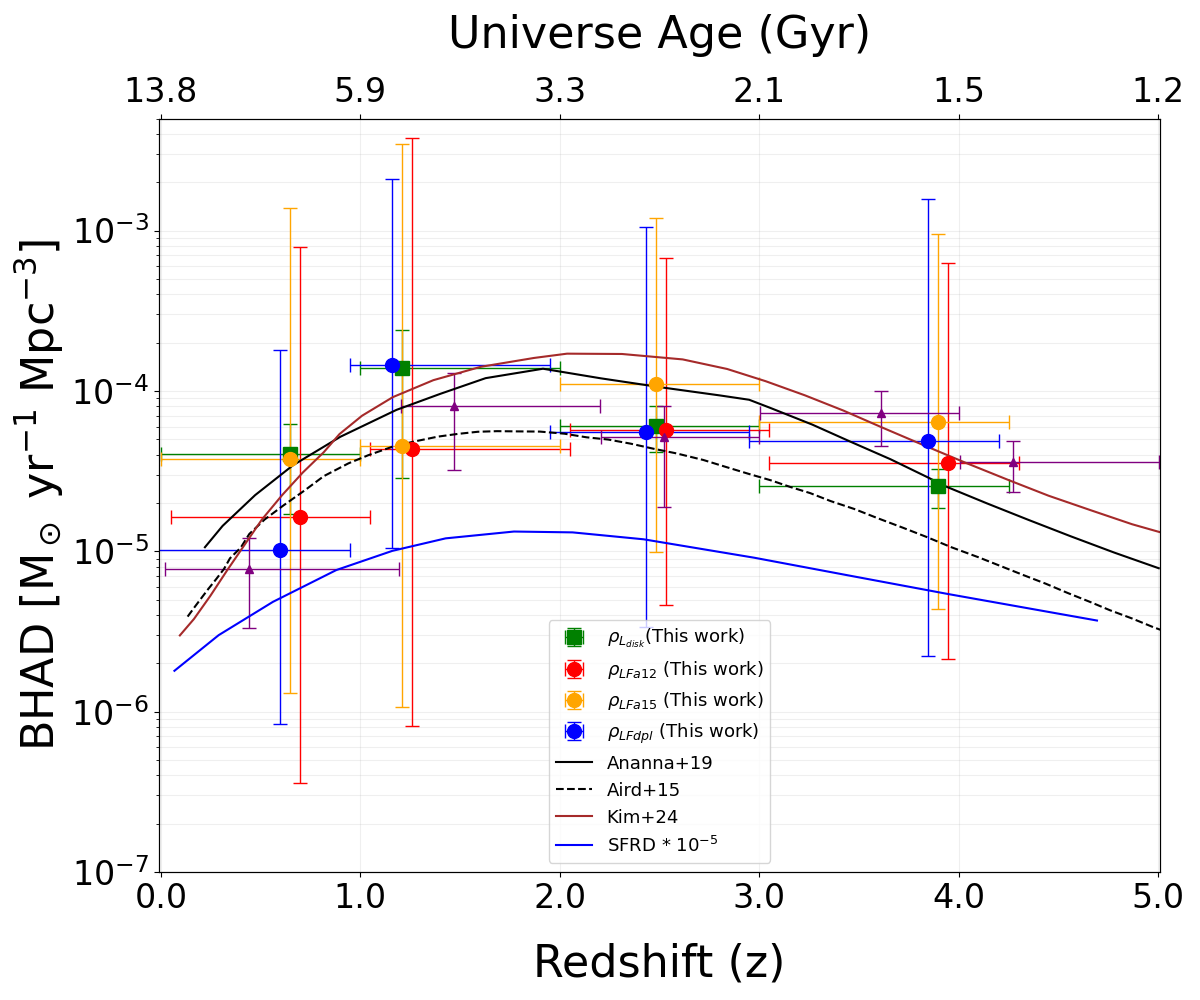}
    \caption{
    Redshift evolution of BHAD. Green squares (labeled as $\rho_{L_{\text{disk}}}$) are derived from the $L_{\text{disk}}$ of the composite and AGN candidates, as explained in Section \ref{2.2}. Their vertical error bars include bootstrap and SED-fitting uncertainties. {Red, yellow, and blue circles (labeled as \( \rho_{\text{LF}} \)) represent values inferred from the AGN LF (Section \ref{3.2}). These are derived using the modified Schechter function with \( \alpha = 1.2 \) and \( \alpha = 1.5 \) and the double power law(dpl), respectively. The x-axis value is the median redshift in the bins, $\rho_{LF_{a12}}$ and $\rho_{LF_{DPL}}$, shifted by 0.05 to display the error bar.} Their vertial error bars indicate the 1 $\sigma$ uncertainty from MCMC. The result from \citet{yang2023ceers} is shown in purple triangles. The horizontal error bars indicate the width of redshift bins.  The black and brown lines denote the two X-ray and one MIR BHAD as reported in previous studies.~\citep{aird2015evolution, ananna2019accretion, kim2024cosmic}, respectively. The blue line represents the star formation rate density (SFRD) from \citet{madau2014cosmic}, scaled down by $10^{-5}$. 
    }
    \label{LF}
\end{figure*}

Subsequently, sum the redshift binned BHAR and divide it by the corresponding comoving volume covered by the MIRI.

\begin{equation}
    \rho_{L_{\text{disk}}} = {\rm BHAD} = \frac{\sum_i \text{BHAR}_i}{V_{max, i}}
\end{equation}

\begin{equation}
V_{max} = \left( V_{c, i}(z) - V_{c, j(z)} \right) \times \frac{9.2\,\text{arcmin}^2}{4\pi}
\label{vmax}
\end{equation}

where $j =  i + 1$, $V_{max}$ is calculated by \texttt{ASTROPY.COSMOLOGY} \citep{2022ApJ...935..167A}.
Hereafter, we use $\rho_{L_{\text{disk}}}$ to represent the BHAD associated with the accretion disk luminosity. Figure \ref{LF} presents the redshift evolution of $\rho_{L_{\text{disk}}}$, in comparison to the result from \citet{yang2023ceers}. 

We note that the $\rho_{L_{\text{disk}}}$ is similar to \citet{yang2023ceers} but reveals a more significant decreasing trend at z $\gtrsim$ 2. This could be due to our stricter criteria for identifying AGNs.

{At \( z \lesssim 1 \), the value of \(\rho_{L_{\text{disk}}}\) is greater than Yang’s work with no overlapping error bars. However, our sample is comparatively smaller, which should result in a lower BHAD. This discrepancy could be due to differences in the redshift binning. \citet{yang2023ceers} excluded low-redshift sources at \( z \lesssim 0.03 \) and defined their first redshift bin as \( z \lesssim 1.2 \). When adopting their redshift binning, our results are slightly higher, but the error bars overlap. Another possible source of discrepancy is the SED fitting process. While \citet{chien2024finding} follows the configuration of \citet{yang2023ceers}, differences in the SED model such as the number of AGN fractions, the dust attenuation model, and redshift determination may affect the derived \( L_{\text{disk}} \) values.}

{Regarding \(\rho_{L_{\text{disk}}}\), the overall declining trend aligns well with previous studies. Although the peak position differs, the magnitude of \(\rho_{L_{\text{disk}}}\) remains consistent with the findings of \citet{yang2023ceers}.} 


In Section \ref{3}, we use our galaxy sample to construct the luminosity function, allowing for a more complete consideration of the unobserved AGNs in the luminosity range from $10^7$ to $10^{13}L_{\odot}$, improving the accuracy of the overall AGN contribution to the BHAD estimation.

\section{Infrared Luminosity Function}\label{3}
\begin{figure*}
    \centering
    \includegraphics[width=1.0\textwidth]{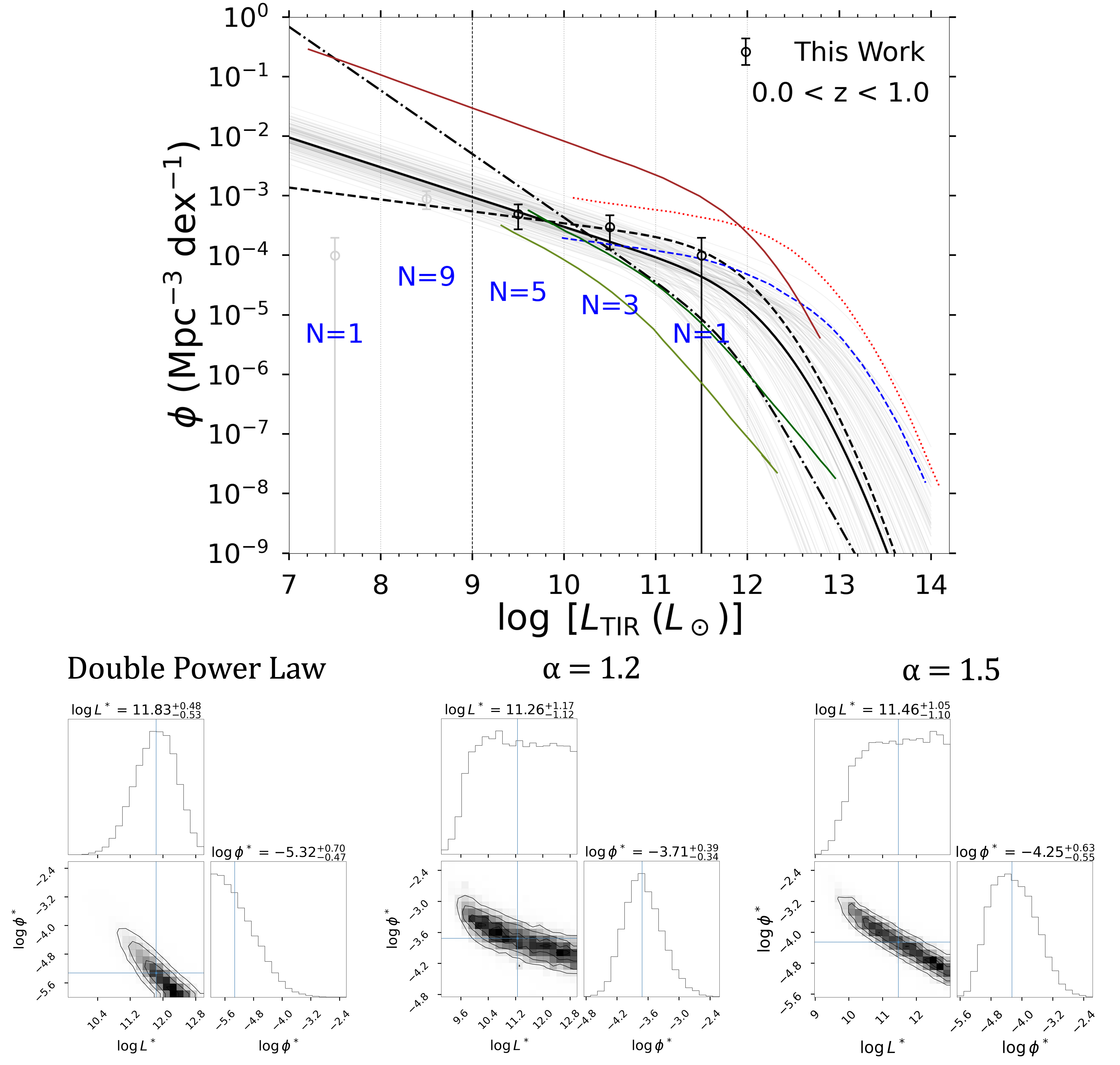}
    \includegraphics[width=0.7\textwidth]{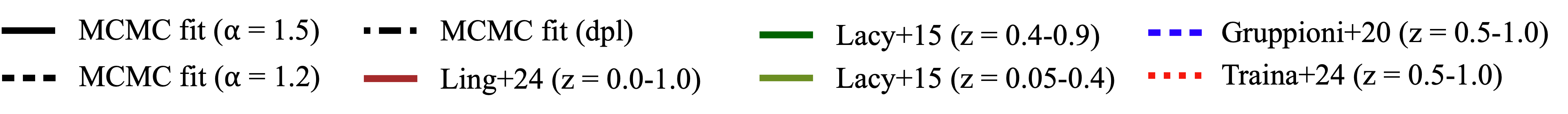}
    \caption{{Above: The rest-frame TIR AGN LF in the redshift range $z = 0-1$. The black line represents the median of the MCMC fit using the modified Schechter function with \( \alpha = 1.5 \), while the gray lines show the 1$\sigma$ uncertainty within {the fit. We} also present the LFs derived from the modified Schechter function with \( \alpha = 1.2 \) (dashed line) and the double power law (dot-dashed line).} A luminosity limit of $L_{\text{TIR}} = 10^9 L_\odot$ is applied for this redshift bin.{Data points that} are not included in the fitting process {are shown by grey open circles.} Galaxy IR LFs \citep{gruppioni2020alpine, ling2024exploring, traina2024a3cosmos} {and AGN IR LF \citep{lacy2015spitzer} from previous studies are shown for comparison.} Below: Corner plot that displays the probability distributions of the parameters obtained from the MCMC analysis. The median values are marked with a blue solid lines, and the 16th and 84th percentiles of the fit parameters are also show in the figure.}
    \label{010}
\end{figure*}
\begin{figure*}
    \centering
    \includegraphics[width=1.0\textwidth]{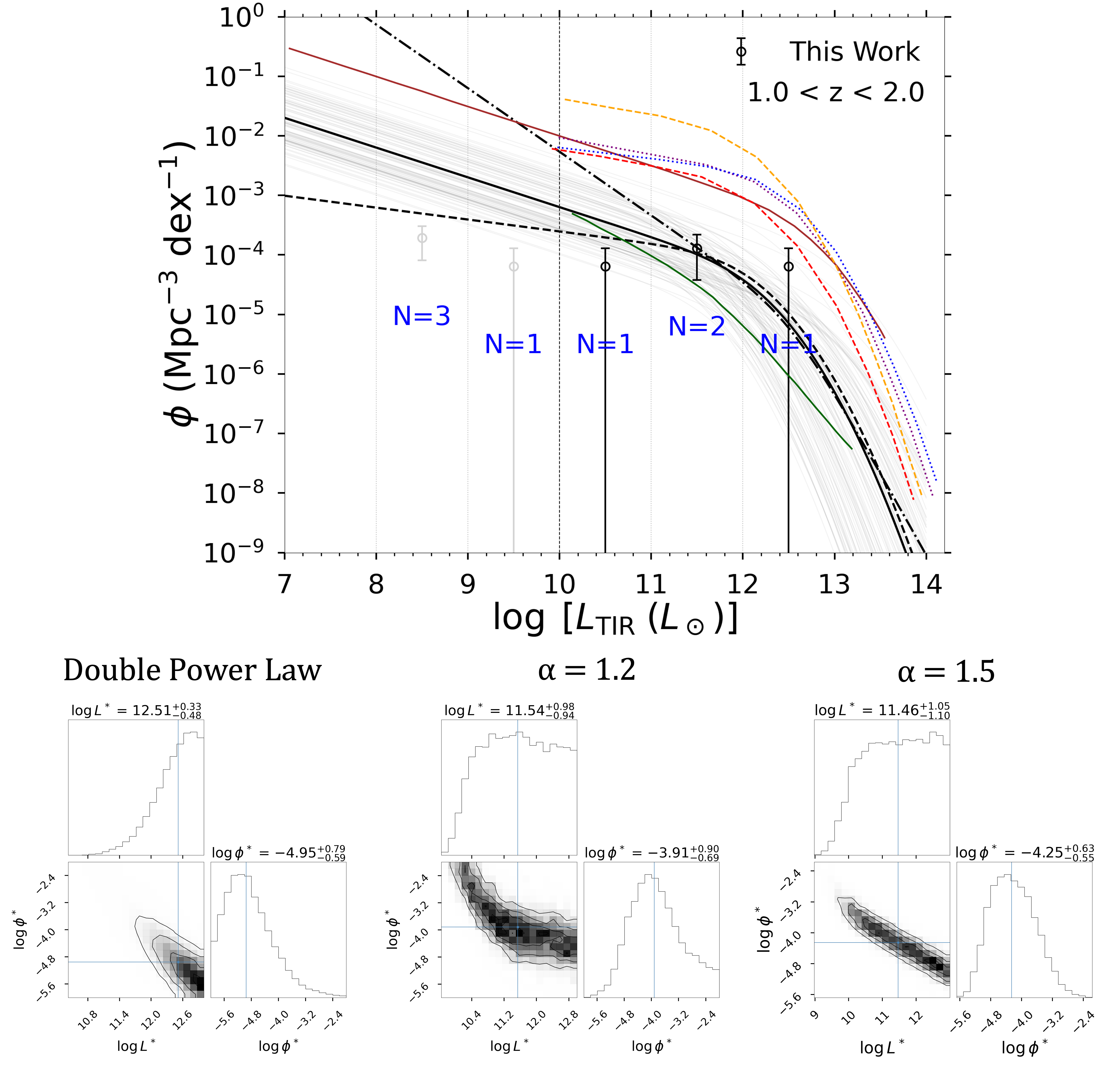}
    \includegraphics[width=1.0\textwidth]{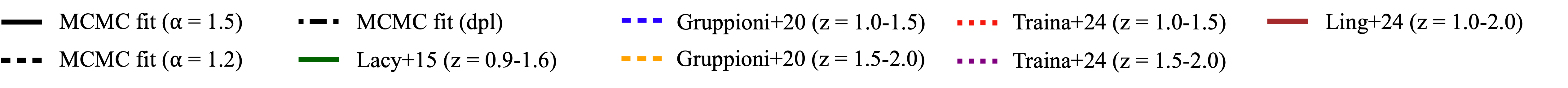}
    \caption{Same as Figure \ref{010}, but for $z = 1-2$. A luminosity limit of $L_{\text{TIR}} = 10^{10} L_\odot$ is applied for this redshift bin and beyond.}
    \label{020}
\end{figure*}
\begin{figure*}
    \centering
    \includegraphics[width=1.0\textwidth]{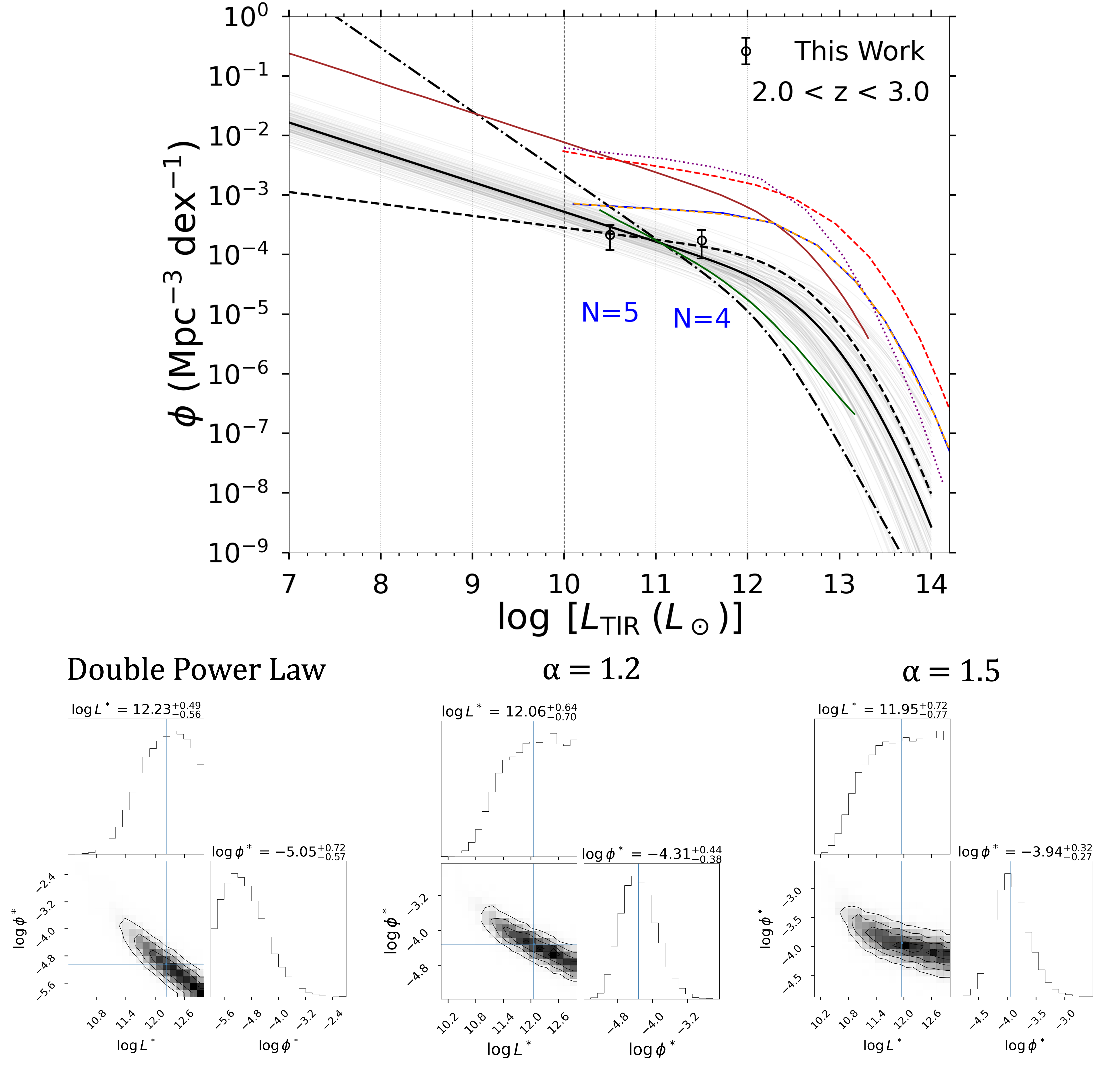}
    \includegraphics[width=1.0\textwidth]{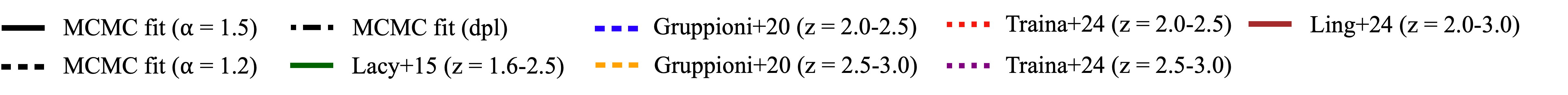}
    \caption{Same as Figure \ref{020}, but for $z = 2-3$.}
    \label{030}
\end{figure*}
\begin{figure*}
    \centering
    \includegraphics[width=1.0\textwidth]{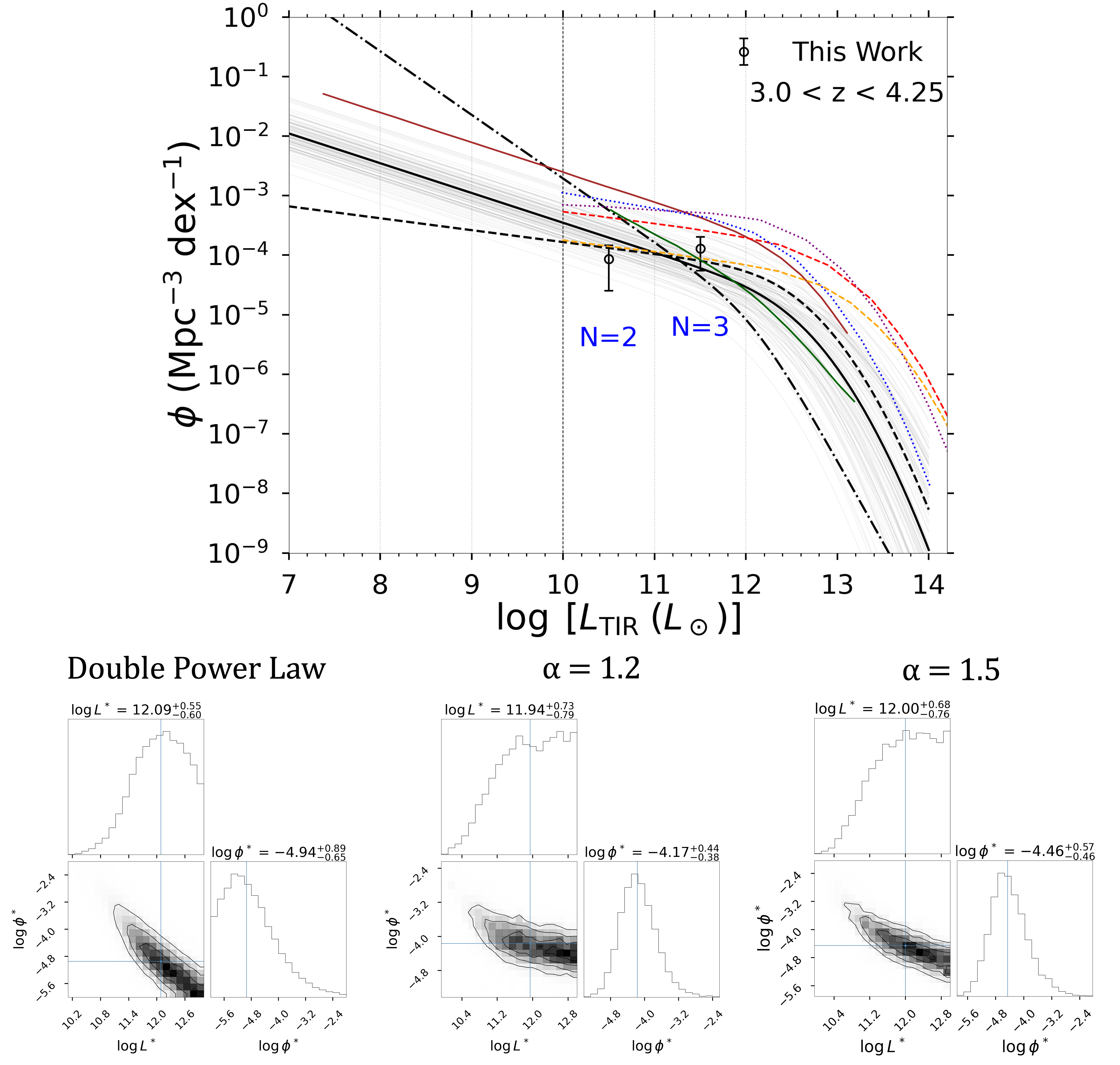}
    \includegraphics[width=1.0\textwidth]{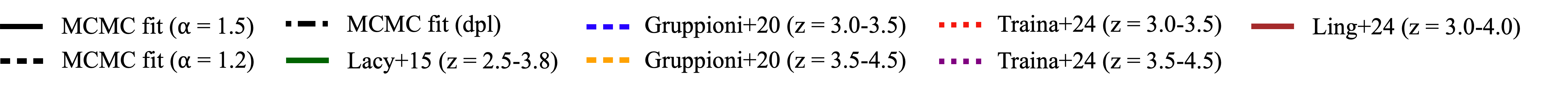}
    \caption{Same as Figure \ref{020}, but for $z = 3-4.25$.}
    \label{040}
\end{figure*}

{This section uses the AGN luminosity function (AGN LF) to determine the BHAD. The AGN LF statistically represents the number density $\phi(L)$ of AGNs as a function of their intrinsic luminosity $L$ within a given volume $V_{max}$, allowing us to account for sources across the full luminosity range. By integrating the AGN LFs multiplied by luminosity, we can more completely capture the contribution of unobserved AGNs to the overall BHAD across cosmic time.}

{To correct for our flux-limited sample, we use the $1/V_{\text{max}}$ method \citep{schmidt1968space}:}
\begin{equation}
    \phi(L) = \frac{1}{\Delta \log L} \sum_i \frac{1}{V_{\text{max},i}}
    \label{1/vmax}
\end{equation}
{where $\Delta \log L = 1.0$ is the luminosity bin width. The maximum observable volume $V_{\text{max}}$ is computed for each source using Equation~\ref{vmax}.}

{The \(V_{\text{max}}\) method estimates the AGN luminosity function by considering the maximum co-moving volume in which each source remains detectable within the survey’s flux limits. For each AGN, the maximum redshift (\(z_{\text{max}}\)) is determined by the survey’s sensitivity, defining its observable range. By correcting for selection biases, this method provides a robust measurement of AGN number density across luminosities and redshifts.}
{To ensure completeness, we impose a limiting luminosity when computing the LFs  to prevent biases in the lowest liminosity bin. The limit are derived from an AGN SED template at the detection threshold for each redshift bin \citep{ling2024exploring}. For redshift bin $0.0 < z < 1.0$, the luminosity limit of $L_{\text{TIR}} = 10^{9} L_\odot$, the other redshift bins are $10^{10} L_\odot$. AGNs below this luminosity threshold are excluded from our LF analysis (see the dashed line in Figures~\ref{010}--\ref{040}).}

Next, we adopt the modified Schechter function \citep{saunders199060} to model the LF:
\begin{equation}
    \phi(L) = \phi^* \left( \frac{L}{L^*} \right)^{1 - \alpha} \exp \left( - \frac{1}{2 \sigma^2} \log_{10}^{2} \left( 1 + \frac{L}{L^*} \right) \right)
    \label{LFeq}
\end{equation}
which behaves as a power law for $L < L^*$ and as a Gaussian for $L > L^*$, where $\alpha$ and $\sigma$ represent the slopes at the faint and bright ends, respectively, $\phi^*$ is the normalization parameter, and $L^*$ is the characteristic luminosity at the knee.

We also adopt a double power law of the form:
    \begin{equation}
    \phi(L) = \phi^* \left[\left(\frac{L}{L^*}\right)^{\gamma_1} + \left(\frac{L}{L^*}\right)^{\gamma_2}\right]^{-1}
    \label{DPL}
    \end{equation}
where $\gamma_1$ and $\gamma_2$ are the slopes at the faint and bright ends, respectively.

{These additional fits allow us to assess systematic uncertainties in the LF modeling and the resulting BHAD estimates.}


\subsection{MCMC Analysis}
{We applied the Markov Chain Monte Carlo (MCMC) method to fit the LF models (Equations~\ref{LFeq}, \ref{DPL}). We used the \texttt{emcee} package \citep{foreman2013emcee} with 100 walkers and 1200 iterations. The fitting prior ranges were set to log($L^*/L_\odot$) = [8, 13] and log($\phi^*/{\rm Mpc}^{-3}\,{\rm dex}^{-1}$) = [-6, -2]. For the modified Schechter function (Equation~\ref{LFeq}) with  we fixed $\alpha = 1.5$ and $\sigma = 0.5$, following \citet{ling2024exploring}. We also fix $\alpha=1.2$ while keeping $\sigma=0.5$, following \citet{gruppioni2013herschel}. This approach allows us to explore the impact of the faint-end slope on the derived BHAD, as $\alpha$
 cannot be reliably constrained with the current data.}

{For double power law (Equation~\ref{DPL}), we fit $L^*$ and $\phi^*$, while $\gamma_1$ and $\gamma_2$ remain fixed, based on the parameters of type 2 AGN LF \citep{lacy2015spitzer}. The prior and range match those of the modified Schechter function.}

{The likelihood function in logarithmic space includes asymmetric Gaussian errors for the observed number density estimates, using data points \(\log y_i\) with uncertainties \(\sigma_{u,i}\) and \(\sigma_{l,i}\).}

\begin{equation}
ln \mathcal{L} = -\frac{1}{2} \sum_{i} \left( \frac{\log y_i - \log y_{\text{model}, i}}{\sigma_i} \right)^2 - \sum_{i} \log \sigma_i
\label{likelihood}
\end{equation}

{where \(\sigma_i\) is determined from the asymmetric errors based on whether \(y_{\text{model}, i} > y_i\) or \(y_{\text{model}, i} < y_i\).}  

{We use an {binned} log-likelihood fit to accurately handle uncertainties in small-number statistics, especially for low-luminosity scenarios where sample size causes significant variations. We also computed the posterior distribution by combining the log-likelihood with priors. To evaluate fit reliability, we used corner plots (Figures~ \ref{010} - \ref{040}) to examine parameter degeneracies and correlations.}

The results of the MCMC analysis for each redshift bin are shown in Figures~\ref{010}--\ref{040}. {The best-fit LF parameters, defined by the median and 16th/84th percentile uncertainties, are summarized in Table~\ref{tab:lf_parameters}. }

\begin{table}[h]
  \centering
  \renewcommand{\arraystretch}{1.2}
  \begin{tabular}{cccc}
    \hline
    $z$ & $L^* (L_{\odot})$ & $\phi^* (\mathrm{Mpc}^{-3} \mathrm{dex}^{-1})$ \\
    \hline
    \multicolumn{3}{c}{\textbf{Modified Schechter Function ($\alpha = 1.2$)}} \\
    \hline
    0.0 - 1.0  & $11.26^{+1.17}_{-1.12}$  & $-3.71^{+0.39}_{-0.34}$  \\
    1.0 -  2.0  & $11.54^{+0.98}_{-0.94}$  & $-3.91^{+0.90}_{-0.69}$  \\
    2.0 - 3.0 & $11.95^{+0.72}_{-0.77}$  & $-3.94^{+0.32}_{-0.27}$  \\
    3.0 - 4.25 & $11.94^{+0.73}_{-0.79}$  & $-4.17^{+0.44}_{-0.38}$  \\
    \hline
    \multicolumn{3}{c}{\textbf{Modified Schechter Function ($\alpha = 1.5$)}} \\
    \hline
    0.0 - 1.0  & $11.46^{+1.05}_{-1.10}$  & $-4.25^{+0.63}_{-0.55}$  \\
    1.0 - 2.0  & $11.66^{+0.91}_{-0.93}$  & $-4.03^{+1.03}_{-0.79}$  \\
    2.0 - 3.0 & $12.06^{+0.64}_{-0.70}$  & $-4.31^{+0.44}_{-0.38}$  \\
    3.0 - 4.25 & $12.00^{+0.68}_{-0.76}$  & $-4.46^{+0.57}_{-0.46}$  \\
    \hline
    \multicolumn{3}{c}{\textbf{Double Power Law}} \\
    \hline
    0.0 - 1.0  & $11.83^{+0.48}_{-0.53}$  & $-5.32^{+0.70}_{-0.47}$  \\
    1.0 - 2.0  & $12.51^{+0.33}_{-0.48}$  & $-4.95^{+0.79}_{-0.59}$  \\
    2.0 - 3.0 & $12.23^{+0.49}_{-0.56}$  & $-5.05^{+0.72}_{-0.57}$  \\
    3.0 - 4.25 & $12.09^{+0.55}_{-0.60}$  & $-4.94^{+0.89}_{-0.65}$  \\
    \hline
  \end{tabular}
  \caption{{MCMC-fit parameters for the AGN LF using the modified Schechter function with $\alpha = 1.2$, $\alpha = 1.5$, and the double power law. The values of $L^*$ and $\phi^*$ are derived from MCMC fitting, with uncertainties representing the 16th and 84th percentiles.}}
  \label{tab:lf_parameters}
\end{table}

LFs from the literature are compared to verify whether our results were consistent with expectations for AGN LFs. 
Since SFGs are excluded, our AGN LFs should generally be lower than the overall galaxy LFs. {We also relate our work to the previous AGN LFs that employ double power laws\citep{lacy2015spitzer}, with which the majority of the AGN LFs in this study intersect.}

{ When comparing the modified Schechter function, we find that the normalization factor \( \phi^* \) for \( \alpha = 1.2 \) is higher than that for \( \alpha = 1.5 \), resulting in a larger integrated region below the LFs. For some modified Schechter function LFs intersect with the galaxy {LFs} \citep{gruppioni2020alpine} within the luminosity range of \( 10^{10} - 10^{11} L_{\odot} \) in Figure \ref{010} and \( 10^{11} - 10^{12} L_{\odot} \) in Figure \ref{040}.  The results may suggest that the expected number density of faint AGNs is higher than previously assumed, or that the number density of faint galaxies should be revised upward. For the double power law, the faint-end intersects with the galaxy LFs from \citet{ling2024exploring} across all redshift bins. These LFs used JWST data to found a higher number density for faint galaxies, with only the double power law LF intersecting their results, this suggests that the faint-end slope of the double power law from \citet{lacy2015spitzer} should be expected to be lower, In Section \ref{dis}, {we will} have more discussion about faint AGN Population.}

Except for the range $1.0 < z < 2.0$, small sample sizes may cause Poisson noise, but MCMC analysis finds an optimal fit within the boundaries, allowing us to infer the Luminosity Density (LD) and BHAD. However, parameter degeneracy may lead to uncertainties that impact their accuracy and reliability.

\subsection{BHAD Evolution}\label{3.2}
The {AGN} luminosity function (LF) {specifies} the average infrared (IR) energy radiated by AGNs within a specific volume. By integrating the LFs {along the luminosity axis} over each redshift bin, we {can precisely derive} BHAD across {cosmic time}:
\begin{equation}
\rho_{\mathrm{LF}} = \frac{\int_{8}^{\infty} (1-\varepsilon) \cdot L \phi(L) \mathrm{d}\log L}{\varepsilon c^2}
\end{equation}
To {measure} the 1$\sigma$ uncertainty (the 16th and 84th percentiles) of $\rho_{LF}$, we use MCMC results presented in Figures \ref{010} - \ref{040}.  The horizontal error bars illustrate the redshift interval of each bin.

The evolution of $\rho_{LF}$ is shown in Figure \ref{LF}, where the results derived from the LF are represented as red dots. "The data indicate that the peak of $\rho_{LF}$ occurs within the redshift range $z \simeq 2 - 3$, highlighting the dominant phase of black hole accretion, when the most rapid growth of supermassive black holes occurred.

\section{Discussion}
\label{dis}

\begin{figure}
    \includegraphics[width=0.8\textwidth]{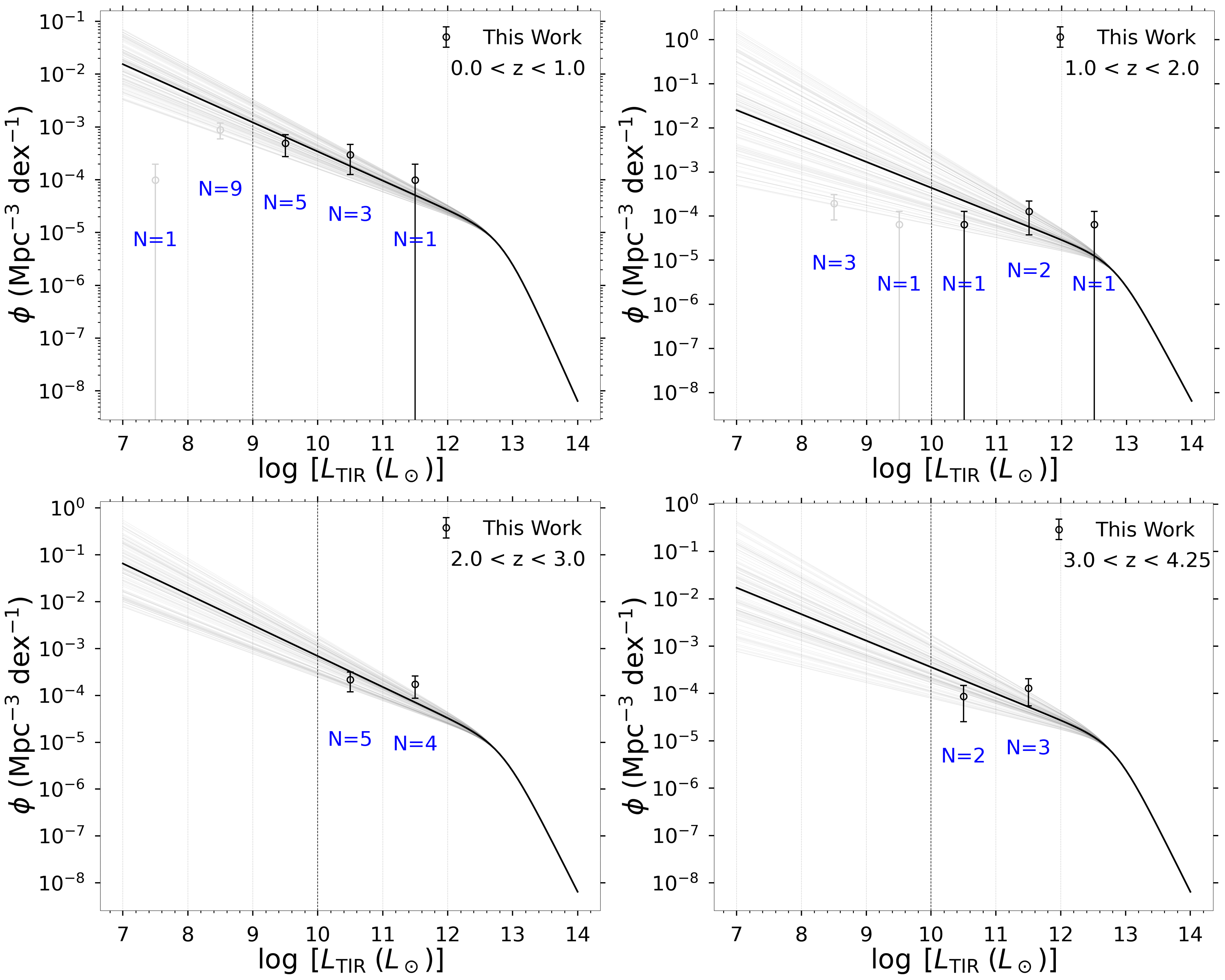}
    \caption{The rest-frame TIR AGN LF, where only the faint-end slope is fitted, is shown across the redshift range $z = 0 - 4.25$. The black line represents the median of the MCMC fit using the double power law with fixed $L^*$, $\phi^*$, and $\gamma_2$, while the gray lines indicate the $1\sigma$ uncertainty. The same luminosity limits as in Figures \ref{010} - \ref{040} are applied. Data points not included in the fitting process are shown as gray open circles.}
    \label{lf_dpl}
\end{figure}

\begin{figure*}
     \centering
    \includegraphics[width=0.5\textwidth]{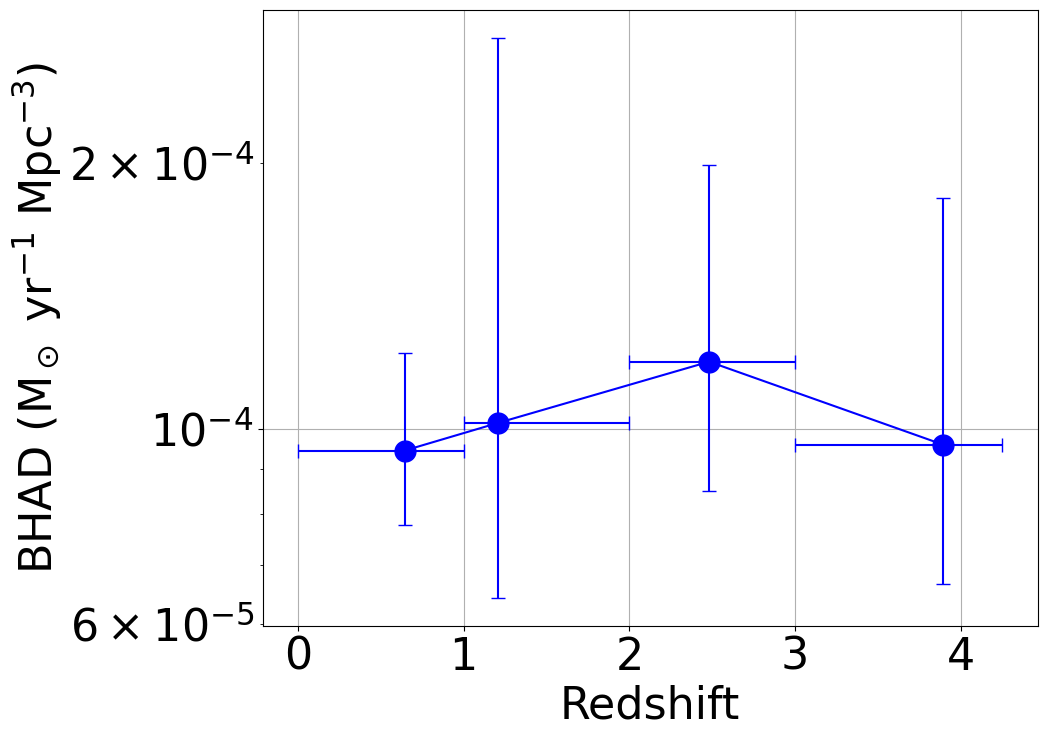}
    \caption{The redshift evolution of BHAD inferred from Figure \ref{lf_dpl}, showing a similar trend to $\rho_{LF_{\text{a12}}}$ and $\rho_{LF_{\text{a15}}}$.}

    \label{bhad_dpl}
\end{figure*}
\begin{table}[h]
  \centering
  \renewcommand{\arraystretch}{1.2}
  \begin{tabular}{cc}
    \hline
    $z$ & \( \gamma_1 \) \\
    \hline
    0.0 - 1.0 & $0.55^{+0.12}_{-0.11}$ \\
    1.0 - 2.0 & $0.59^{+0.32}_{-0.30}$  \\
    2.0 - 3.0 & $0.66^{+0.17}_{-0.17}$ \\
    3.0 - 4.25 & $0.56^{+0.24}_{-0.24}$ \\
    \hline
  \end{tabular}
  \caption{The faint-end slope for LFs determined from the median of the MCMC fitting using the double power law. }
  \label{para_gamma}
\end{table}

{We have presented four estimations of the BHAD, $\rho_{L_{disk}}$ (Section \ref{2.2}) and three types of $\rho_{LF}$: $\rho_{LF_{a12}}$, $\rho_{LF_{a15}}$ and $\rho_{LF_{dpl}}$   (Section \ref{3.2}), based on AGN accretion disk luminosity ($L_{disk}$) and LFs, respectively. The LF method provides a more complete analysis of potentially unobserved sources, making $\rho_{LF}$ better represent the overall contribution of AGNs for BHAD, even though one to two order of magnitude uncertainties remain, likely due to Poisson noise.} In Figures \ref{010} and \ref{020}, some luminosity bins contain only a single source, which may affect the fitting results. Despite these uncertainties, the LFs provide useful insights into the population of AGNs across the luminosity range from $10^7$ to $10^{13}L_{\odot}$.

{Due to JWST data limitations, we fixed the faint-end slope \(\alpha\) of modified Schechter function and \(\gamma_1\) of double power law, potentially affecting \(L^*\). Compared to the AGN LFs of \citet{lacy2015spitzer}, our \(L^*\) is higher, indicating the discrepency may be due to the differences in LF form. We explore this discrepancy by comparing BHAD evolution trends below.}

{Next, we discuss the trends of $\rho_{L_{\text{disk}}}$ and $\rho_{LF}$. These estimates exhibit a similar increasing trend, peaking before declining. Previous studies also report different peak at different redshifts. The X-ray studies \citep{aird2015evolution} finds a peak at $z \simeq 1.5$, while \citet{ananna2019accretion} reports $z \simeq 2$, consistent with $\rho_{L_{\text{disk}}}$ and $\rho_{LF_{\text{dpl}}}$ peaking at $1 \lesssim z \lesssim 2$. The IR study \citep{kim2024cosmic} identifies a peak at $z \simeq 2.3$, consisting with $\rho_{LF_{\text{a12}}}$ and $\rho_{LF_{\text{a15}}}$, which peak at $2 \lesssim z \lesssim 3$.}

{The discrepancy between peak redshifts and BHAD from $L_{\text{disk}}$ can be attributed to $\rho_{LF}$, derived by integrating LFs, which includes contributions from fainter AGNs possibly missed in direct $L_{\text{disk}}$ measurements. This suggests accretion disk luminosity might not fully represent the accretion process at the period, But the  $\rho_{L_{\text{disk}}}$ indicate the lower limits of the BHAD. }

{The discrepancy in \(\rho_{LF}\) may due to the choice of the LF form. Since we adopt a fixed slope from previous studies \citep{gruppioni2013herschel, lacy2015spitzer, ling2024exploring}, most of our sources have luminosities below $10^{12}L_{\odot}$, unlike pre-JWST studies.This suggests that the assumed faint-end slope may differ from the actual number density distribution of AGNs, potentially leading to discrepancies in the derived BHAD. The results indicate that AGN LF models require further consideration of the faint population, as the choice of LF model can influence the inferred BHAD evolution. This contrasts with the conclusion of \citet{vslaus2023xxl}, found that modified Schechter function and double power law have consistent results for the LDDE model.}

{Therefore, we focus on the faint population JWST observed. We fit only the faint-end slope $\gamma_1 $ of the double power law, following the other parameters fixed following the type 2 AGN LF from \citet{lacy2015spitzer}. Figure \ref{lf_dpl} is the AGN LFs, the parameter are summarized in Table \ref{para_gamma}. To compare our work with \citet{lacy2015spitzer}, their \(\gamma\) is \(\gtrsim 1\), while our value is smaller, indicating a lower expected number density for the faint AGN population. These results suggest a stronger constraint on the faint-end slope of AGN LFs. Figure \ref{bhad_dpl} presents the results of BHAD evolution infer from the Figure \ref{lf_dpl}. The trend remains consistent with the BHAD derive from the modified Schechter function LFs, exhibiting a peak at \( z = 2 - 3 \) followed by a decline. This improvement enhances our understanding of the BHAD evolution of the faint AGN population is following the \(\rho_{LFa15}\), \(\rho_{LFa12}\) and previous studies. In the following discussion, \(\rho_{LF}\) will be used to represent the trends of both. }

Then, we compare our results with the star formation rate density (SFRD) described by \citet{madau2014cosmic}, which peaks at redshift $z \simeq 2$. Previous research suggests that accretion processes may delay star formation events, indicating a relationship between these two phenomena \citep{hickox2014black}. This relationship is also observed in AGNs at $1.5 \lesssim z \lesssim 2.5$ \citep{rodighiero2015relationship}.

\begin{figure*}
     \centering
    \includegraphics[width=0.5\textwidth]{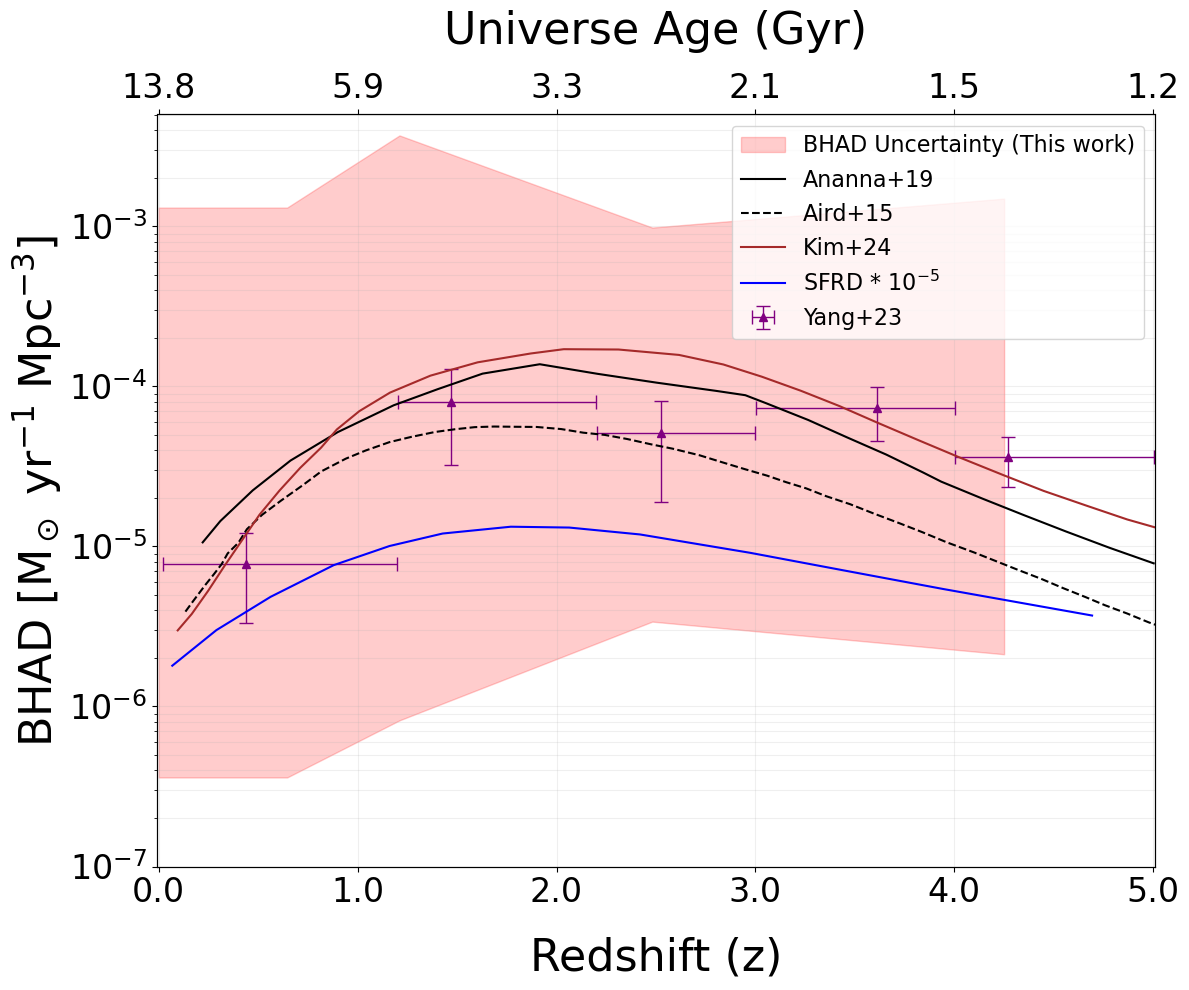}
    \caption{Summary plot of BHAD across redshift, showing the conservative upper and lower limits derived from different assumptions in this work (shaded red band). For comparison, \citet{yang2023ceers} (purple triangles).\citet{aird2015evolution, ananna2019accretion, kim2024cosmic} and (SFRD) from \citet{madau2014cosmic}, scaled down by $10^{-5}$ are ploted.}

    \label{bhad}
\end{figure*}

{Figure \ref{bhad} summarizes the BHAD range across our assumptions at each redshift.} It raises a fundamental question: {AGN} activity precede or follow the evolution of cosmic star formation? As noted by \citet{yang2023ceers},a scenario where {AGNs} appear first would challenge existing black hole formation theories, which typically posit that star formation initiates galaxy evolution in the early universe \citep[e.g.,][]{habouzit2021supermassive, zhang2023trinity}.
Our present study, based on early JWST observations, provides an initial exploration of {AGN} evolution. However, the limited sample size precludes statistically significant conclusions. The forthcoming availability of larger, statistically robust JWST datasets will be crucial for accurately quantifying these discrepancies and ultimately elucidating the intricate interplay between black hole and galaxy co-evolution.

If so, as pointed out by \citet{yang2023ceers}, this result might challenge the current black hole formation theories, as current simulations predict that the star-formation occurs first in the early universe \citep[e.g.,][]{habouzit2021supermassive, zhang2023trinity}. New theoretical models might be needed to explain this observational result. {Although our results do not lead to a definitive conclusion, this study serves as an exploratory analysis using JWST data to investigate BHAD evolution. Future JWST statistical data will be crucial for quantifying these differences and understanding black hole co-evolution. As uncertainties decrease, if the BHAD/SFRD ratio does not rise, theoretical predictions and observations will remain consistent.}

Finally, we compare our work with \citet{yang2023ceers}. They inferred the BHAD at $z \gtrsim 3$ is approximately 0.5 dex higher compared to X-ray study results. $\rho_{LF}$ also shows a similar result. { However, due to overlapping error bars, we can only suggest that the value may be higher than X-ray studies. More candidates are needed to confirm this result with greater confidence.} This trend may be due to contributions from faint and highly obscured sources that are only detectable in MIR bands \citep{yang2023ceers, eckart2009comparison}.  

At high redshifts, the neutral gas surrounding AGN is likely denser and more abundant compared to lower redshift environments. This results in significantly higher column densities that can effectively block X-ray radiation, making it difficult to detect these AGN through X-ray observations, while the mid-IR observations can \citep[e.g.,][]{gilli2022supermassive}.
Still, \citet{yang2023ceers} should be considered as lower limits of the BHAD, because they did not account for faint galaxies below the detection limit, while our LF method ($\rho_{LF}$) does.

At $z \lesssim 3$, $\rho_{LF}$ is comparable to the BHAD obtained from X-ray studies \citep{ananna2019accretion} and consistent with the MIR study \citep{kim2024cosmic}. This trend indicates that using LF provides a more complete analysis of the missing sources, but the error bars still require more data to constrain. 

\section{Conclusions}
We have presented the black hole accretion history associated with obscured AGNs identified through the JWST CEERS mid-IR survey. We developed the luminosity functions to deduce the $\rho_{LF}$ corresponding to these selected candidates. {Our results are summarized as follows:}

\begin{enumerate}
\item We applied a simple method to estimate $\rho_{L_{disk}}$ similar to that described in \citet{yang2023ceers}. Our results show a more significant trend compared to pre-JWST studies.

\item {We construct AGN IR lFs from JWST data, testing models such as the modified Schechter function with (\(\alpha = 1.2\), \(\alpha = 1.5\)) and a double power law. Our LFs included type 2 AGNs up to two orders of magnitude fainter than those in pre-JWST surveys. Compared to previous galaxy and AGN LFs, we find a similar order and $L^*$ , though additional data are needed to enhance accuracy and mitigate Poisson noise.}

\item {By integrating the LFs and multiplying by the luminosity, we present the evolution of BHAD $\rho_{LF}$ with three forms, including candidates as faint as $10^7 L_{\odot}$. The estimates show a similar increasing trend, peaking at $z \sim 2$ for $\rho_{L_{\text{disk}}}$ and $\rho_{LF_{\text{dpl}}}$, and at $z \sim 3$ for $\rho_{LF_{\text{a15}}}$ and $\rho_{LF_{\text{a12}}}$. The discrepancy may arise from accretion disk luminosity not fully capturing the accretion process or differences in functional forms and assumptions and the choice effect of LF form. The results suggest that AGN LF models require more consideration of the faint AGN population. We also examine the faint-end slope of the double power law. The BHAD derived from this model is consistent with that from the modified Schechter function LFs and previous studies, supporting the BHAD evolution of the faint AGN population.}

\item  {{Whether the AGN activity precedes or follows the evolution of cosmic star formation is still a challenge for existing black hole formation theories.} Our present study, based on early JWST observations, provides an initial exploration of {AGN} evolution. However, the limited sample size precludes statistically significant conclusions. The larger JWST datasets in the future will be critical for determining this relationship and understanding the co-evolution of black holes and galaxies.}

\end{enumerate}

This study highlights that JWST's high sensitivity in mid-IR offers a new opportunity to discover a larger population of faint, obscured AGNs.  Such mid-IR studies are important because the obscured fraction may increase with redshift \citep{hickox2018obscured}, and sources at redshift $z \sim 2$ typically show weak X-ray detections \citep{stern2015x}. A deep JWST mid-IR survey is thus required to address these issues of detecting faint, obscured AGNs at high redshifts. Obtaining further observational data from JWST to reduce the uncertainty is to explore accretion events in the early universe more accurately.

\begin{acknowledgement}

The authors sincerely appreciate the anonymous referee for constructive comments that significantly enhanced the paper. The authors acknowledge the Taiwan Astronomical Observatory Alliance (TAOvA) and its summer student internship program for partial financial support of this research. TG and Tetsuya Hashimoto acknowledge the support of the National Science and Technology Council of Taiwan through grants 113-2112-M-007-006, 113-2927-I-007-501, and 113-2123-M-001-008.
This work is based on observations made with the NASA/E-SA/CSA JWST. The data were obtained from the Mikulski Archive for Space Telescopes at the Space Telescope Science Institute, operated by the Association of Universities for Research in Astronomy, Inc., under NASA contract NAS 5-03127 for JWST. These observations are associated with program JWST-ERS01345. The authors also acknowledge the CEERS team for developing their observing program with a zero-exclusive-access period.
This work is based on observations taken by the CANDELS Multi-Cycle Treasury Program with the NASA/ESA Hubble Space Telescope (HST), which is also operated by the Association of Universities for Research in Astronomy, Inc., under NASA contract NAS 5-26555.
This work utilized high-performance computing facilities operated by the Center for Informatics and Computation in Astronomy (CICA) at National Tsing Hua University, funded by the Ministry of Education of Taiwan, the National Science and Technology Council of Taiwan, and the National Tsing Hua University.
\end{acknowledgement}

\printendnotes
\printbibliography
\end{document}